\documentclass[journal,10pt]{IEEEtran}

\usepackage{cite}
\usepackage{amsmath,amssymb,amsfonts}
\usepackage{graphicx}
\usepackage{textcomp}
\usepackage{xcolor}
\usepackage{multirow}
\usepackage{array}
\usepackage{algorithm,algpseudocode}
\usepackage{titlesec}
\titlespacing*{\subsection}{0pt}{6pt}{3pt}

\usepackage[
top    = 0.7 in,
bottom = 1.03 in,
left   = 0.65 in,
right  = 0.65 in]{geometry}

\begin{document}

\title{
Task-oriented Framework for Communication-Efficient Federated Learning: From Isolated Optimization to Holistic Synergy
}

\author{Fuqiang~Pan,
        Yan~Liu,
        Erwu~Liu,
        and~Arumugam~Nallanathan,~\IEEEmembership{Fellow,~IEEE}
\thanks{This work was supported in part by the Intergovernmental International Science and Technology Innovation Cooperation of National Key Research \& Development Program of China (2024YFE0197400), the National Natural Science Foundation of China (62401395). \textit{(Corresponding author: Yan Liu.)}}
\thanks{F. Pan, Y. Liu and E. Liu are with the Department of Information and Communication Engineering, Tongji University, Shanghai 201804, China (e-mail: \{2431926, yanliu2022, erwuliu\}@tongji.edu.cn).}
\thanks{A. Nallanathan is with the School of Electronic Engineering and Computer Science, Queen Mary University of London, London, UK (e-mail: a.nallanathan@qmul.ac.uk).}
}

\maketitle

\begin{abstract}
Communication bottlenecks remain a primary obstacle to the large-scale deployment of federated learning (FL).
This article proposes a comprehensive framework for building communication-efficient FL, founded on three fundamental pillars: model compression, client selection, and resource allocation.
We first survey state-of-the-art techniques for each pillar, specifically elucidating how quantization, pruning, and low-rank approximation reduce payloads; how intelligent client schedulers exploit heterogeneity; and how emerging communication paradigms such as Integrated Sensing and Communication (ISAC) and Over-the-Air Computation (AirComp) redefine bandwidth and energy utilization.
Subsequently, these insights are unified through a task-oriented design philosophy that couples strategy selection with cross-layer, multi-objective optimization.
To validate the proposed framework, we present an autonomous driving case study with two complementary experiments: a task-oriented client scheduling strategy that improves object detection accuracy under the same communication time budget, and a joint quantization-bandwidth optimization that further reduces total training time under dynamic networks.
Together, the experiments demonstrate the advantages of holistic task-oriented design for real-world FL deployment.

\end{abstract}


\section{Introduction}
The rapid development of Machine Learning (ML) has driven significant progress across scientific and technological domains.
Traditional ML frameworks often require the centralized collection, processing, and training of personal data, which increases privacy and security risks.
For example, banks collaborating to develop ML-based fraud detection models may need to share transaction data to improve model accuracy, inadvertently exposing customers' financial details.
To address this issue, Federated Learning (FL) has emerged as a more suitable, privacy-preserving, decentralized ML paradigm.
FL enables multiple devices (clients) to collaboratively learn a shared model by exchanging locally computed model updates (e.g., gradients or parameters) instead of raw data.
This approach is particularly suitable for scenarios where data sharing is restricted due to policy, security, or network capacity constraints.

Traditional FL systems rely on iterative, high-precision exchanges of model parameters over wireless channels to achieve collaborative training.
This process faces several communication challenges that significantly affect the system's efficiency.
First, limited and heterogeneous spectrum resources struggle to accommodate the transmission of large-scale model parameters.
Second, constrained and varied energy resources limit the sustainability of frequent communication and computation.
Finally, communication delays and unstable connectivity may lead to training interruptions or gradient loss.
These intertwined communication barriers are particularly pronounced in large-scale FL applications.
As model sizes grow or the number of participating clients increases, communication overhead escalates exponentially, leading to higher energy consumption and prolonged training convergence time.

In this context, traditional FL assumptions of ``lossless transmission, full client participation, and uniform resource allocation'' are no longer applicable.
Inefficient communication is widely recognized as a critical impediment to the scalable deployment of FL frameworks in production environments.
To alleviate communication bottlenecks, reduce client-side computational burdens, and ensure model convergence performance, researchers have proposed various efficiency optimization strategies tailored to resource-constrained scenarios.
This paper focuses on three principal and complementary approaches: Model compression, Client selection, and Resource allocation.

\begin{itemize}
    \item \textbf{Model compression} reduces the volume of transmitted data while maintaining model performance through techniques such as quantization, pruning, and low-rank approximation.
    \item \textbf{Client selection} coordinates client participation based on device characteristics (e.g., data quality, resource availability) to enhance FL performance in heterogeneous environments.
    \item \textbf{Resource allocation} addresses the bandwidth and energy bottlenecks of individual devices by optimizing the distribution of communication and computation resources, often incorporating cutting-edge wireless technologies.
\end{itemize}

While these three pillars have motivated extensive research, existing works predominantly
  optimize individual pillars in isolation~\cite{jiang2022model,wolfrath2022haccs,cao2021optimized}, achieving efficiency gains within their respective domains.
However, this isolated approach neglects inter-pillar coupling: compression noise affects the value of high-quality clients, while bandwidth budget determines the feasible compression level.
Recent works~\cite{liu2022training,zhang2022multi} begin to integrate multiple pillars, yet they remain focused on intermediate communication metrics rather than task-level performance such as model accuracy or convergence.

In contrast, this paper proposes a task-oriented framework that achieves holistic synergy among the three pillars.
Our contributions include:
\begin{itemize}
    \item \textbf{Systematic Framework Establishment:} This paper proposes a comprehensive optimization framework for communication-efficient FL.
    This framework systematically analyzes the principles and breakthroughs of three pillar technologies.
    Notably, our discussion on resource allocation incorporates emerging communication paradigms, highlighting their potential to reshape FL transmission.
    \item \textbf{Task-Oriented Design Philosophy:} This paper introduces a task-oriented design philosophy that directly aligns strategy selection with end performance objectives, supported by a dynamic cross-layer collaborative optimization approach that coordinates the three pillars for global efficiency.
    \item \textbf{Case Study Validation:}
    To demonstrate the practical efficacy of our framework, we design and evaluate a task-oriented strategy for the Internet of Vehicles (IoV), a domain characterized by severe heterogeneity in devices, networks, and data.
    Simulation experiments conducted on the CARLA platform validate the necessity and effectiveness of our proposed strategy, demonstrating superior performance in a realistic autonomous driving application.
\end{itemize}

\section{Three Pillars of Communication-Efficient FL}
\begin{figure}[htbp]
    \centering
    \includegraphics[width=\columnwidth]{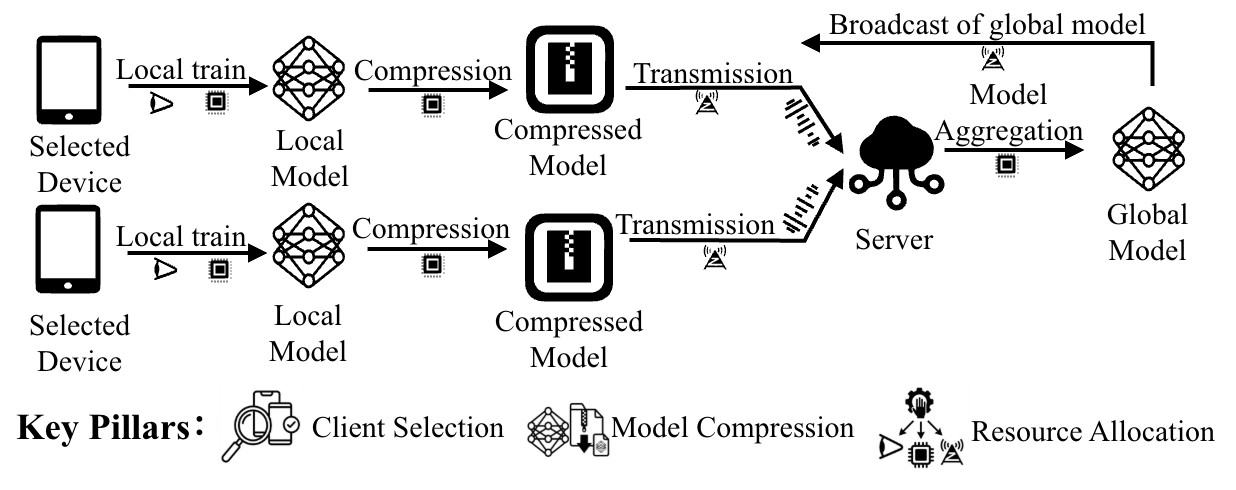}
    \caption{Workflow of communication-efficient FL illustrating the three key pillars: client selection, model compression, and resource allocation.}
    \label{fig:three_pillars}
\end{figure}

Improving the energy and spectrum efficiency is a key challenge for enabling large-scale distributed FL, especially in resource-constrained scenarios.
To address this challenge, researchers have proposed multi-dimensional optimization techniques, primarily categorized into three pillars: model compression, client selection, and resource allocation, as shown in Fig. \ref{fig:three_pillars}.
In this section, we analyze each technology individually, covering its principles, functions, application scenarios, and representative works.

\subsection{Model Compression}
In resource-constrained FL, reducing communication overhead in each round is a key direction for enhancing system efficiency.
Traditional FL algorithms transmit full-precision model parameters or gradient updates, leading to low spectrum efficiency and high communication latency.
Consequently, model compression techniques have been introduced to reduce the amount of transmitted model data.
This subsection discusses three primary methods: quantization, pruning, and low-rank approximation.

\subsubsection{Quantization}

Quantization significantly reduces communication overhead by converting the floating-point parameters into lower-precision representations, thereby decreasing the number of bits required for each transmission.
However, quantization noise leads to inaccurate client updates, causing the global model to deviate from its optimal convergence path, which not only increases the number of required training rounds but also degrades final accuracy.

The Quantized Stochastic Gradient Descent (QSGD) algorithm \cite{alistarh2017qsgd} is a foundational approach, offering a trade-off between communication cost and convergence variance.
It allows computing nodes to adjust the bit-width of gradients to reduce transmission costs at the expense of increased variance.
While effective, QSGD applies uniform quantization across all clients, which may be suboptimal under heterogeneous client conditions.

To address this limitation, advanced strategies have been proposed to optimize quantization bit allocation based on specific client characteristics like channel quality or data significance.
For example, work \cite{lan2023quantization} formulates the allocation as a non-convex integer programming problem aimed at minimizing an upper bound on the convergence error under a total bit budget.
A nearly optimal solution is derived using the Karush-Kuhn-Tucker (KKT) conditions and linear search.
Numerical results demonstrate that this scheme significantly improves the performance and robustness of FL.
Other emerging directions include quantization-aware training \cite{sun2022lazily} and mechanisms to compensate for quantization errors in subsequent iterations \cite{qu2025fedqclip}.
\subsubsection{Pruning}
Pruning is another technique that reduces the number of model parameters by removing less important neurons or connections, thus reducing communication costs in FL.
The ``lottery ticket hypothesis'' suggests that within large neural networks, optimized subnetworks exist that can achieve the same performance as the original network, which forms the foundation for pruning techniques.
However, typical pruning techniques only consider centralized datasets and static models, which may not suit the dynamic FL environment.

PruneFL \cite{jiang2022model} is a dynamic pruning method specifically designed for FL.
It performs initial pruning on selected clients and adaptively adjusts the model size and fine-tunes remaining parameters throughout the training process.
Experiments on four different datasets on real edge devices (Raspberry Pi) show that PruneFL effectively identifies the relative importance of different network layers and adaptively adjusts the model complexity.
PruneFL can reduce transmitted data volume and improve communication efficiency while maintaining comparable model accuracy.

\subsubsection{Low-Rank Approximation}
In FL, model parameters or gradients can be represented as matrices.
Approximating these matrices with lower-rank versions that have fewer parameters could reduce consumption costs, which is referred to as the low-rank approximation technique.
Specifically, this is typically achieved by minimizing the difference between the original matrix and its approximated version, where closed-form solutions are often obtained by Singular Value Decomposition (SVD).
This method faces similar challenges as quantization, i.e., the trade-off between reduced resource cost and preserved model accuracy, as low-rank models often have limited expressive capacity.
To address this challenge, FedPara \cite{hyeon2021fedpara} employs a low-rank Hadamard product parameterization, which eliminates low-rank constraints and achieves higher model accuracy compared to traditional low-rank approximation methods.
More recent work also extends low-rank approximation to federated split learning, further reducing the upload payload by transmitting only compact low-rank factors of the activation maps \cite{ao2026federated}.
Simulations show that FedPara can achieve comparable performance to original models while reducing resource costs by 3 to 10 times.

\subsection{Client Selection}
Traditional FL usually assumes that all clients participate in each training round.
In practice, limited spectrum prevents the central server from communicating with all clients simultaneously, necessitating the selection of a subset of clients per round.
Early FL algorithms usually use random and unbiased client selection methods, which fail to optimize efficiency under client heterogeneity in terms of network conditions, computational capabilities, and data distributions.
Consequently, a variety of client selection algorithms have emerged to optimize FL performance, which are mainly categorized into data-based, resource-based, and hybrid approaches.
The following will discuss these three types of strategies in detail and explore how rational client selection can enhance FL efficiency.

\subsubsection{Data-Based}
The data-based algorithm selects clients based on the quality and representativeness of their local data.
The core idea is to accelerate global model convergence by prioritizing clients whose data distributions align well with the global objective or whose data is of higher quality.
Key challenges of this approach lie in effectively evaluating data quality and identifying the most representative clients.
Various metrics are used to assess data quality, including local dataset size, the $\ell_2$ norm of model updates, the local loss values, and so on.

One of the representative examples is ShapleyFL \cite{singhal2024greedy}, which treats FL as a cooperative game and calculates the Shapley value of each client based on its contribution to the global model update.
Then according to the Shapley values, the selection probability of each client is dynamically adjusted.
ShapleyFL can perform well in scenarios where certain nodes have data distributions significantly deviating from the global model or where there are malicious nodes.

The standard calculation of Shapley values suffers from high computational complexity and poor scalability.
To address this, the GreedyFed algorithm \cite{singhal2024greedy} has been proposed, which integrates a fast approximation algorithm, GTG-Shapley, with a purely greedy strategy to simplify the calculation.
Simulation results show that GreedyFed effectively reduces communication costs and accelerates convergence while ensuring model robustness.

\subsubsection{Resource-Based}
The resource-based strategy prioritizes clients based on their available resources, such as computational power, network spectrum, energy level, and other conditions.
Selecting well-resourced clients can significantly reduce per-round latency and communication costs, thereby accelerating overall training.

One of the typical resource-based algorithms is TiFL \cite{chai2020tifl}, where clients are grouped into clusters based on their computational performance.
In each round, a cluster is selected according to two factors: the average loss and the sampling frequency in each cluster.
The former measures the model's performance within the cluster, thereby accelerating model improvement, while the latter ensures that no cluster is over-sampled or under-sampled, maintaining a fair distribution of training opportunities across all clusters.
This balances workload and ensures efficient resource utilization.

\subsubsection{Hybrid (Data- and Resource-Based)}
The hybrid method combines the advantages of both data-based and resource-based approaches, with the core idea being to flexibly choose clients based on their data value and available resources.

The Heterogeneity-Aware Clustered Client Selection (HACCS) is a typical hybrid algorithm \cite{wolfrath2022haccs}, with a hierarchical structure similar to TiFL.
First, clients generate local data summaries (i.e., label distribution histograms) and send them to the central server.
Then, the server groups similar clients into clusters and applies weighted random sampling within these clusters, balancing average loss and normalized latency.
By prioritizing lower-latency clients from each cluster, HACCS reduces per-round training time while maintaining data diversity.
Simulation results show that HACCS can reduce model training time by 18\%-38\% compared to TiFL.

Another typical algorithm, FedMarl \cite{zhang2022multi}, utilizes Multi-Agent Reinforcement Learning (MARL) to select clients.
Specifically, by designing a weighted reward function, FedMarl can simultaneously consider model accuracy, processing latency (the sum of training and communication delays), and communication cost.
Experiments show that FedMarl can achieve better accuracy than full client participation while reducing latency by 41\% and communication costs by 55\%, with good adaptability to diverse FL settings.

\subsection{Resource Allocation}
Efficient resource allocation is one of the core challenges in FL, as the training process consumes substantial communication and computational resources.
Traditional optimizations focus on tuning parameters, such as transmission power, channel spectrum, and device CPU/GPU computing frequencies.
The objective is typically to minimize metrics like per-round training time or total energy consumption under given system constraints.

However, as the scale of FL systems continues to grow, these conventional approaches are revealing significant bottlenecks in both spectrum and energy efficiency, struggling to meet the demands of large-scale, highly dynamic scenarios.
To overcome these limitations, recent research turns to emerging communication paradigms that natively integrate computation with the wireless transmission.
This subsection focuses on two emerging technologies: Integrated Sensing and Communication (ISAC) and Over-the-Air Computation (AirComp).

\subsubsection{Integrated Sensing and Communication (ISAC)}
ISAC enables wireless signals to perform dual functions, namely data communication and environmental sensing, using shared hardware and spectrum resources.
This integration enhances spectrum and energy efficiency through resource reusing, and can dynamically allocate resources between communication and sensing tasks to achieve more flexible system optimization.

Exploring the potential of ISAC for FL involves jointly optimizing Sensing, Computation, and Communication (SC$^2$) resources.
For example, in an FL system where ISAC devices act as clients, one can formulate a joint optimization problem.
A key insight is that sensing quality improves with transmission power but plateaus after reaching a certain threshold.
Based on this observation, an optimized algorithm has been proposed in \cite{liu2022training} optimizing the allocation of sensing and communication power, along with client batch sizes, to maximize FL convergence speed.
This joint resource allocation problem is formulated as a single-variable optimization problem solved via one-dimensional broad search.
Simulation results confirm the superiority of the joint SC$^2$ resource allocation scheme in FL performance compared to baseline approaches.

\subsubsection{Over-the-Air Computation (AirComp)}
AirComp exploits the waveform superposition property of wireless channels to perform functional computation (e.g., summation) directly over the air.
This technology fundamentally changes the communication resource consumption pattern of model aggregation in FL.
It is particularly suitable for FL with limited spectrum resources, as it allows for model aggregation directly during wireless transmission, thus improving bandwidth utilization, reducing latency, and mitigating privacy leakage risks.
Recent extensions further adapt this principle to semi-federated learning (semi-FL) settings with computing-heterogeneous IoT devices, jointly designing communication and over-the-air aggregation to enhance scalability \cite{ni2025joint}.

A primary challenge is aggregation error caused by channel fading and noise.
To reduce this impact, a transmission power optimization algorithm has been proposed \cite{cao2021optimized}.
It formulates a convex optimization problem via Lagrangian duality to maximize convergence speed (i.e., minimizing the optimality gap) by optimally allocating transmit power among clients.
Simulation results show that this algorithm significantly accelerates the convergence compared to fixed-power or minimum mean square error (MSE) strategies.

Beyond traditional FL scenarios, recent advances in federated foundation models (FedFM) bring new challenges to communication-efficient FL at unprecedented scales~\cite{ren2026federated}.
Foundation models such as GPT-3 (175 billion parameters) are orders of magnitude larger than traditional FL models (typically $<$10 million parameters), fundamentally altering the three pillars' roles.
Model compression shifts from optional to mandatory, as parameter-efficient methods (LoRA, adapters) that update <1\% of parameters become necessary.
Client selection must consider computational capacity alongside data quality, as most edge devices cannot host billion-parameter models.
Resource allocation faces bandwidth demands that shift from megabytes to gigabytes per round.
These challenges make isolated pillar optimization insufficient, motivating the holistic, task-oriented framework presented in the following section.

\section{Comprehensive Framework for Communication-Efficient FL}
\begin{figure}[htbp]
    \centering
    \includegraphics[width=\columnwidth]{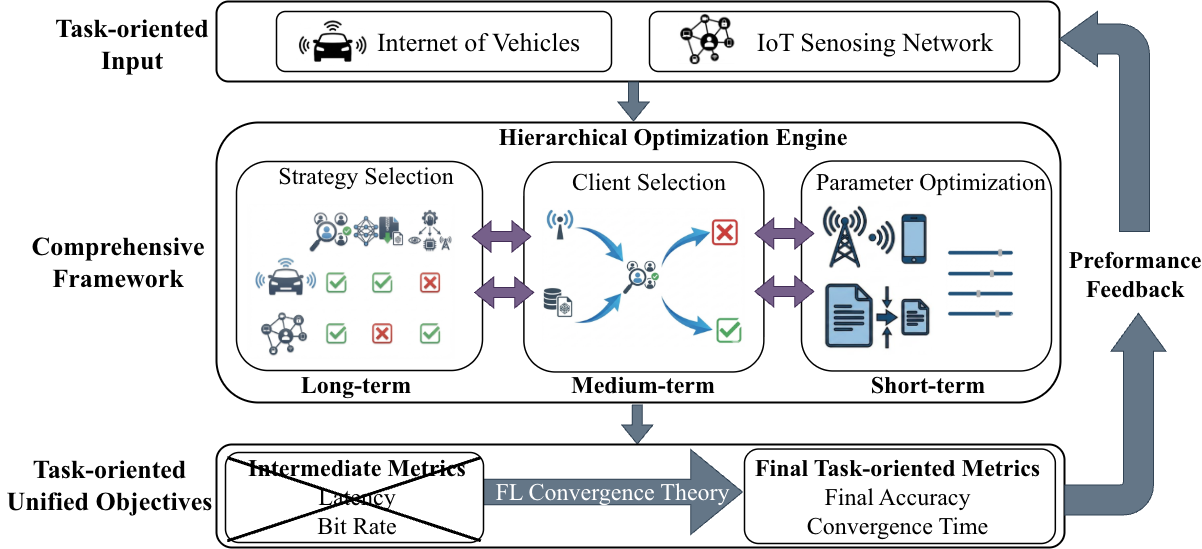}
    \caption{A comprehensive framework for communication-efficient FL, which integrates hierarchical optimization across three timescales (long-term strategy selection, medium-term client selection, and short-term parameter tuning) with task-oriented objectives, enabling continuous adaptation through performance feedback.}
    \label{fig:framework}
\end{figure}

While the three pillars individually advance FL efficiency, current research predominantly pursues their optimization in isolation.
This siloed approach fails to capture their interdependencies, potentially leading to suboptimal system performance.
For example, an aggressive compression strategy that performs well alone might degrade performance when combined with a specific client selection or resource allocation strategy.
Practical FL deployments face concurrent challenges: communication overhead reduction, device heterogeneity mitigation, resource efficiency enhancement, and so on.
These challenges necessitate a comprehensive, co-designed optimization approach.
In this paper, we propose a comprehensive framework to coordinate these three pillars by addressing key questions on technology selection, objective setting, and collaborative optimization, as shown in Fig. \ref{fig:framework}.

\subsection{Task-Oriented Strategy Selection}
FL tasks vary widely, imposing fundamentally different demands on the system.
Therefore, strategy selection should be guided by specific task characteristics and requirements.
For instance, in autonomous driving scenarios, vehicles exhibit diverse hardware, experience varying network conditions, and collect non-IID data from different road environments.
These factors collectively present multiple heterogeneous challenges to the system.
A hybrid client selection strategy (considering both data quality and resource status) is ideal to balance efficiency and performance.
In contrast, severely resource-constrained Internet of Things (IoT) devices face challenges in different dimensions.
These battery-powered devices with limited compute capacity require aggressive model pruning or low-rank decomposition to reduce communication overhead, while leveraging sophisticated resource allocation strategies to maximize energy and spectrum efficiency, ensuring sustained participation without battery exhaustion.

This task-driven strategy selection is far from a static mapping.
Systems must possess real-time sensing and dynamic adaptation capabilities: increasing compression rates when network is congested to reduce data transmission volume, prioritizing clients with high-quality data during critical convergence phases to accelerate training, and adjusting communication frequency during energy-scarce periods to extend device lifetime.

\subsection{Task-Oriented Unified Optimization}
For a long time, FL research has focused on intermediate metrics such as communication latency, training rounds, or transmitted bits, neglecting their complex relationship with their ultimate task-level performance.
A 50\% reduction in communication volume does not guarantee a proportional improvement in system performance; excessive compression can even lead to a significant degradation in model accuracy, requiring more rounds to converge.

Truly meaningful optimization should directly target task-level metrics (e.g., final model accuracy, total time to achieve a target accuracy), which are the performance aspects that end-users genuinely care about.
In recent years, an increasing number of studies have begun to leverage FL convergence theory to guide system design \cite{liu2022training}.
Rather than relying on empirical parameter tuning, these works delve into the mathematical impact of factors like compression noise, client sampling bias, and communication errors on convergence speed.
Recent research shows that this approach not only generally improves task performance, but also enhances the interpretability of decision-making and provides a unified optimization objective for integrating multiple strategies.

\subsection{Multi-Objective Collaborative Framework}
To achieve further gains, the three pillars must be jointly optimized.
However, complex interdependencies and constraints exist: model compression affects the reliance on high-quality clients, client selection influences data/resource distribution, and resource allocation constrains feasible compression/selection options.
Therefore, FL system design must adopt a holistic, system-level perspective, avoiding local optima for individual modules in favor of global performance maximization.

A practical approach is to adopt a hierarchical optimization strategy that operates across different timescales: the long-term level determines the base strategy combination based on historical experience and task profile; the medium-term level adjusts client selection according to network conditions and device availability; while the short-term level optimizes compression parameters and communication resources in real-time to adapt to instantaneous changes.
This hierarchical architecture reduces complexity through timescale separation and achieves global coordination, ensuring the system can rapidly respond to environmental changes while maintaining long-term performance stability.

\section{Case Study in IoV}
FL has emerged as a promising paradigm for collaborative intelligence among IoV devices due to its decentralized and privacy-preserving nature.
However, practical deployment in IoV faces multifaceted heterogeneity challenges: device heterogeneity (varying hardware), network heterogeneity (dynamic wireless channels for mobile vehicles), and data heterogeneity (non-IID data from different driving perspectives).
This scenario serves as an ideal testbed for applying our proposed task-oriented design philosophy, where client selection must directly optimize for the end-to-end task objective (e.g., total training time to achieve target accuracy) rather than intermediate communication metrics.

To validate this philosophy, we conduct two complementary experiments.
The first focuses on the \emph{client selection} pillar and minimizes total training time as the scheduling objective; the second jointly tunes the \emph{model compression} and \emph{resource allocation} pillars under the same objective to examine cross-pillar coordination gains.

\subsection{Experiment 1: Task-Oriented Client Selection}
We propose a task-oriented client selection strategy that minimizes the total training time required to achieve the target accuracy.
Through convergence analysis, we derive a closed-form expression for the total training time that correlates the objective with two key client-specific variables: 1) gradient importance, quantified by the norm of local model updates and representing each client's data contribution to global model improvement; and 2) communication resources, determined by the client's instantaneous and time-varying channel quality and available bandwidth, which relates to model upload latency.
Based on this theoretical foundation, we formulate the problem of minimizing total training time as a convex optimization problem, whose solution provides dynamic scheduling probabilities for each vehicle in each communication round.

To validate the effectiveness of the proposed scheduling strategy, we conduct simulations on the CARLA high-fidelity autonomous driving platform.
The collaborative task involves training a YOLOv5 object detection model, a well-established network architecture widely deployed for real-time detection in autonomous driving applications.
We simulate a fleet of seven autonomous vehicles, each serving as an FL client, operating within the urban and suburban environments provided by CARLA's ``Town03'' and ``Town05'' maps.
Since the distance between each vehicle and the edge server continually changes while driving, the wireless channel conditions remain heterogeneous and time-varying throughout the training process.
Each vehicle uses only its onboard front-facing camera data for local training, thereby naturally generating a non-IID data distribution that mimics real-world conditions.

\begin{figure}[h]
    \centering
    \includegraphics[width=0.7\columnwidth]{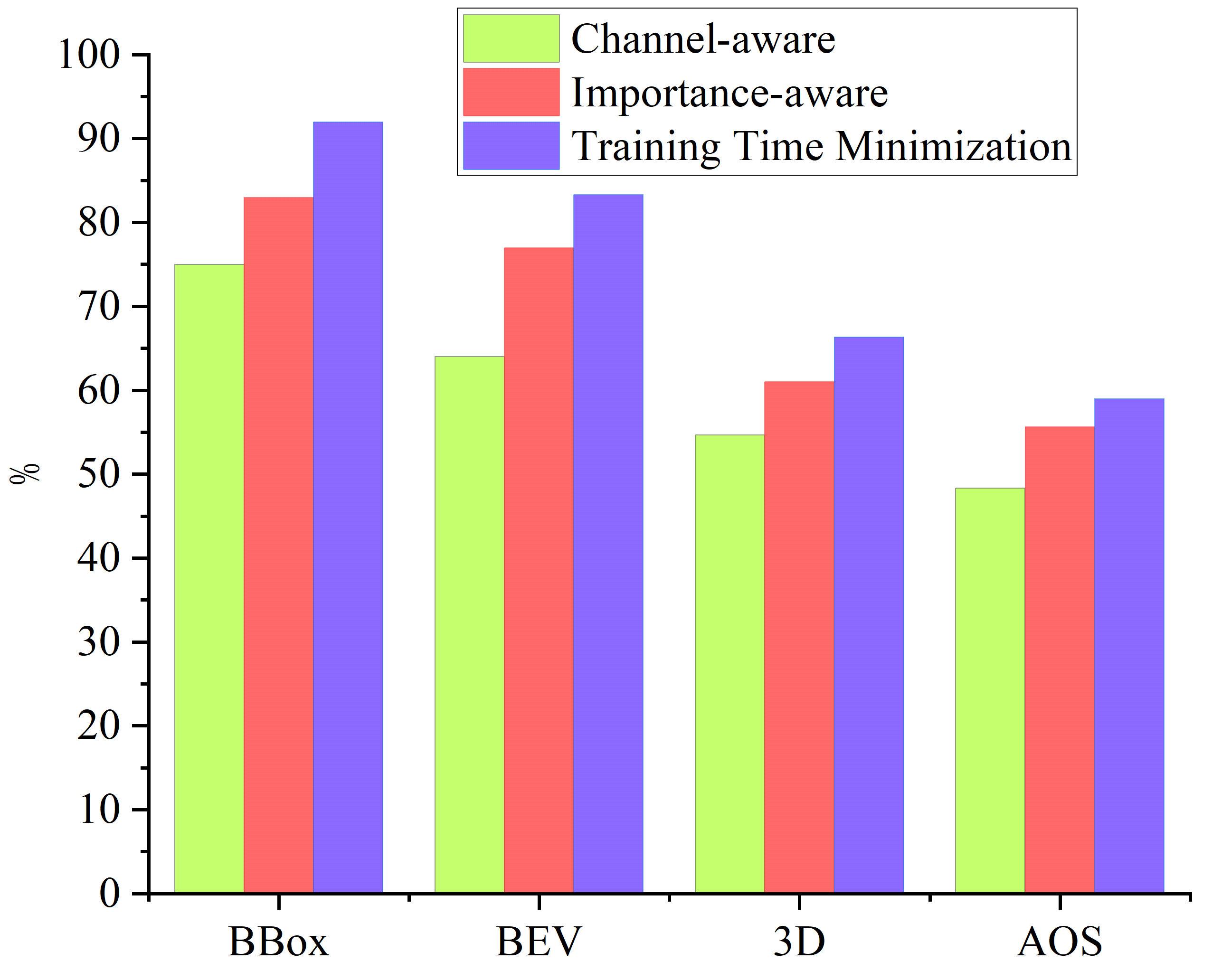}
    \caption{Performance comparison of channel-aware, importance-aware, and training-time-minimization strategies on four object detection metrics.}
    \label{fig:g2}
\end{figure}

Under the same communication time budget, we evaluate performance using four object detection metrics: 2D Bounding Box (BBox), Bird's Eye View (BEV), 3D Bounding Box (3D), and Average Orientation Similarity (AOS) Average Precision.
As shown in Fig. \ref{fig:g2}, our training time minimization (TTM) strategy consistently achieves the highest accuracy across all metrics compared to channel-aware or importance-aware baselines.
This demonstrates that holistic optimization for the end-to-end task objective more effectively navigates the trade-off between data utility and communication efficiency.

\subsection{Experiment 2: Joint Compression and Resource Allocation}
\begin{figure}[htbp]
    \centering
    \includegraphics[width=0.8\columnwidth]{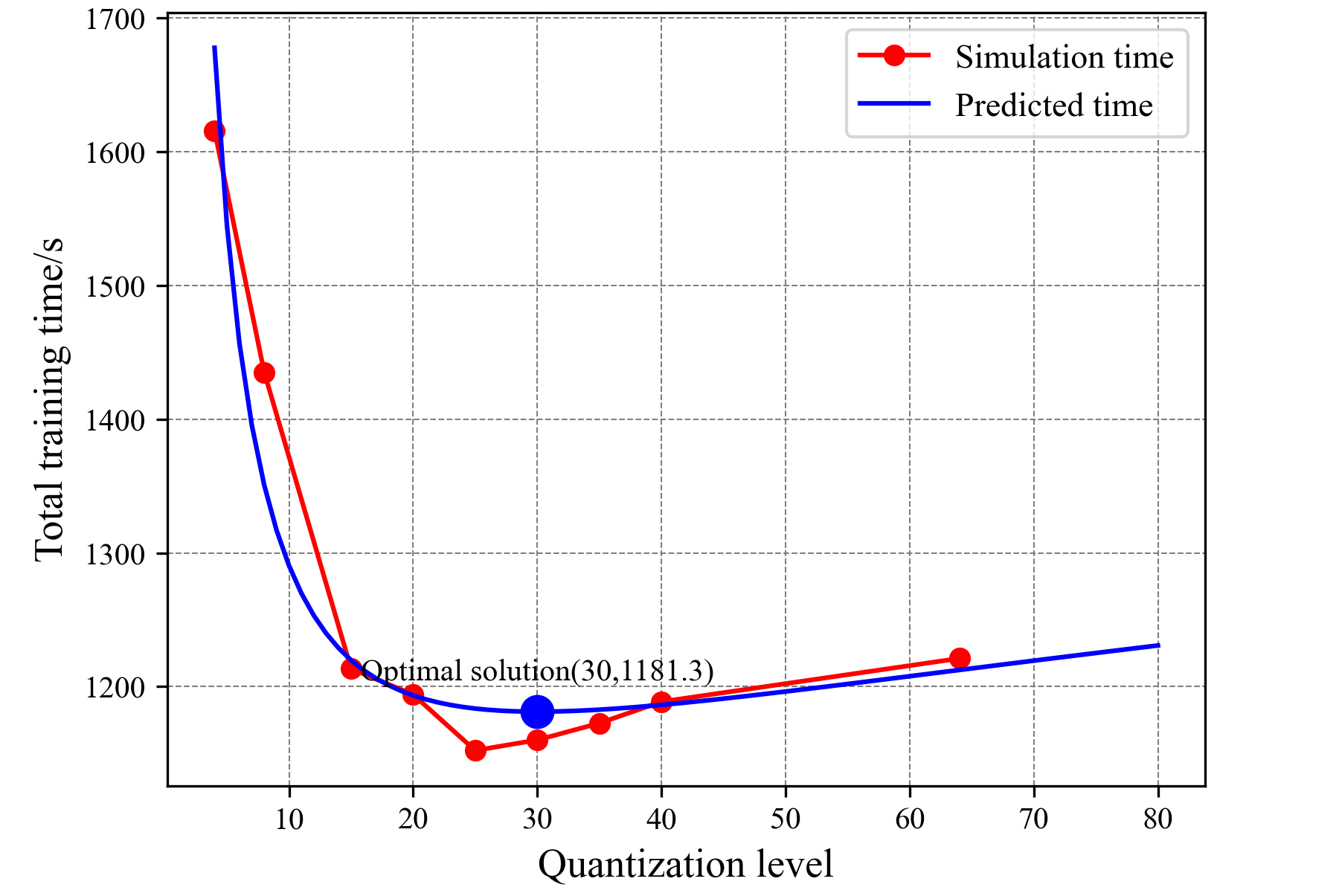}
    \caption{Total training time under different quantization levels.}
    \label{fig:single_column_image1}
\end{figure}

Building on Experiment 1, we now examine the multi-pillar coordination axis by jointly tuning quantization and bandwidth allocation under the same task-level objective.
We first characterize how quantization levels trade off convergence quality against per-round transmission cost, then formulate the total training time minimization as a convex problem under a bandwidth budget.
Simulation results in Fig. \ref{fig:single_column_image1} show that our analytical model accurately fits the relationship between quantization level and total FL training time.

\begin{figure}[htbp]
    \centering
    \includegraphics[width=0.8\columnwidth]{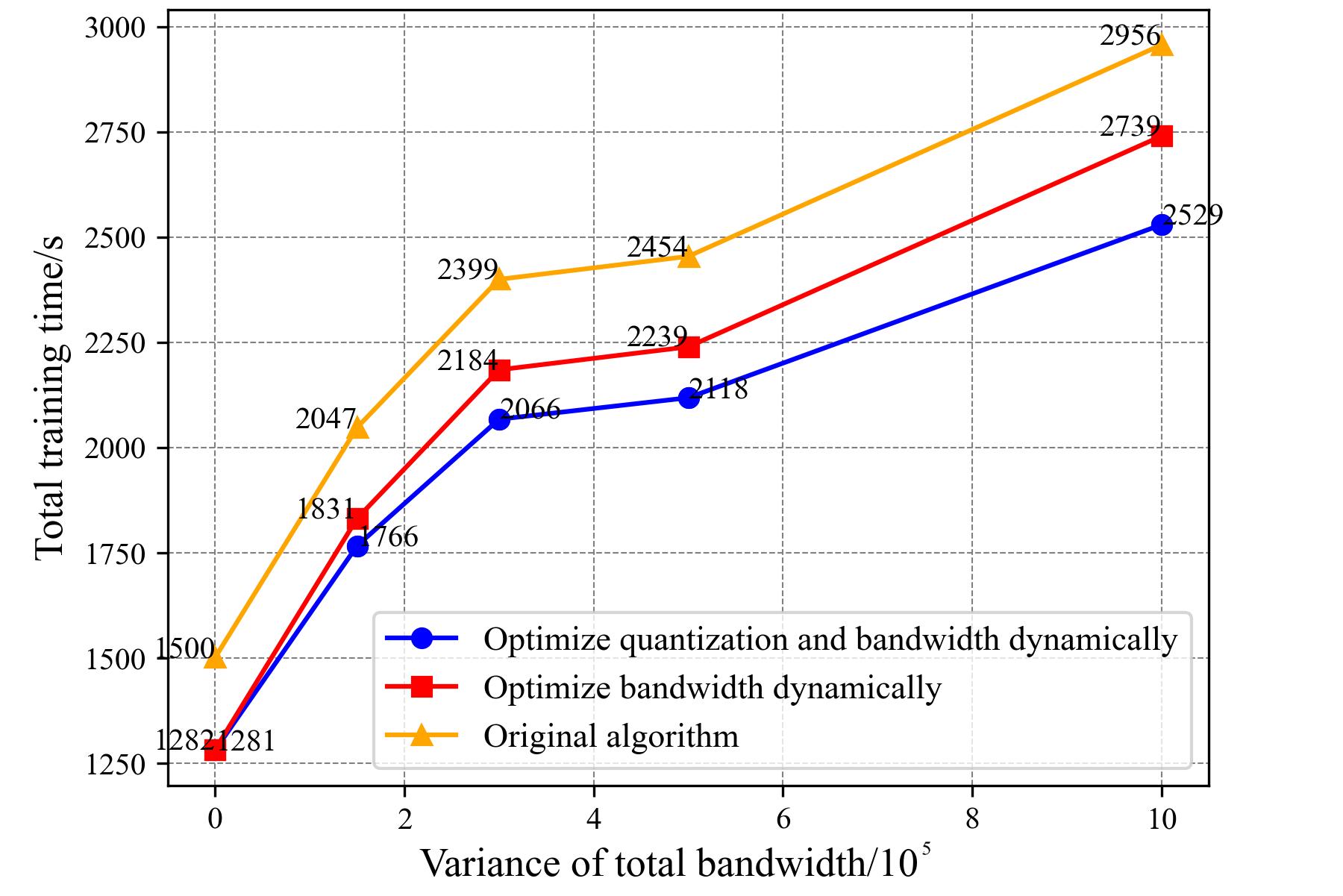}
    \caption{Total training time of different algorithms under different bandwidth variances.}
    \label{fig:single_column_image2}
\end{figure}

The total training time predicted by the above algorithm may be inaccurate in dynamic networks, where device and channel conditions vary over time.
To address this problem, we propose an improved algorithm called Joint Dynamic Optimization of Quantization and Bandwidth (JDOQB), which focuses on optimizing each communication round individually to maximize the reduction of the optimization gap per unit time, under the assumption of predictable time-varying channels.
As shown in Fig. \ref{fig:single_column_image2}, results in different network environments show that the JDOQB consistently outperforms the original algorithm, especially in more drastic time-varying networks.

Collectively, the two experiments validate three design choices of our framework: (i) selecting the right per-pillar strategy (TTM in Experiment 1), (ii) targeting end-to-end task-level objectives over intermediate communication metrics, and (iii) jointly optimizing interdependent pillars (JDOQB in Experiment 2).
These results confirm the necessity of holistic, task-oriented FL design.


\section{Conclusion}
This paper proposes a comprehensive optimization framework to address the critical communication challenges in FL.
Built upon three fundamental pillars (model compression, client selection, and resource allocation), the framework provides analysis, integration, and a task-oriented design philosophy.
The proposed comprehensive design framework moves beyond isolated technological improvements toward task-oriented unified optimization that directly targets end-to-end performance metrics.
Two complementary experiments validate the framework in realistic autonomous driving deployments: a task-oriented scheduling strategy improves object detection accuracy, and a joint quantization-bandwidth optimization further
reduces total training time under dynamic channels.
These results demonstrate that unified, task-driven design offers a viable path for transitioning FL from research prototypes to real-world applications.
However, real-world deployment still faces practical challenges including asynchronous straggler mitigation, gradient inversion attacks, and system scalability for massive IoT deployments.
Future work will explore reinforcement-learning-driven cross-pillar coordination to handle asynchronous stragglers, investigate privacy-preserving mechanisms against gradient inversion attacks, and extend the framework to massive heterogeneous IoT deployments.
\bibliographystyle{ieeetr}
\bibliography{ref}

\end{document}